\documentclass[twocolumn,pra,amssymb,amsmath]{revtex4}
\usepackage{graphicx,epsfig,subfigure}

\begin{document}

\title{Higher-order Brunnian structures and possible physical realizations}
\author{N.~A.\ Baas}
\affiliation{Department of Mathematical Sciences, NTNU, NO-7491
  Trondheim, Norway}
\author{D.~V.\ Fedorov, A.~S.\ Jensen, K.\ Riisager, A.~G.\ Volosniev,
  and N.~T.\ Zinner}
\affiliation{Department of Physics and Astronomy, Aarhus University, 
  DK-8000 Aarhus C, Denmark}

\begin{abstract}
  We consider few-body bound state systems and provide precise
  definitions of Borromean and Brunnian systems. The initial concepts
  are more than a hundred years old and originated in mathematical
  knot-theory as purely geometric considerations. About thirty years
  ago they were generalized and applied to the binding of systems in
  nature.  It now appears that recent generalization to higher order
  Brunnian structures may potentially be realized as laboratory made
  or naturally occurring systems.  With the binding energy as measure,
  we discuss possibilities of physical realization in nuclei, cold
  atoms, and condensed matter systems. Appearance is not
  excluded. However, both the form and the strengths of the
  interactions must be rather special. The most promising subfields
  for present searches would be in cold atoms because of external
  control of effective interactions, or perhaps in condensed-matter
  systems with non-local interactions.  In nuclei, it would only be by
  sheer luck due to a lack of tunability.
\end{abstract}

\maketitle

\section{Motivation}
Tying knots has from ancient times been a very practical skill to
master. Two circular closed strings (rings) can be tied together and
only removed from each other after cutting one of them.  Such
arrangements are called linked geometries.  With more than two rings
numerous possibilities arise for distinctly different topological
structures.  If one part cannot be removed from another part without
untying or cutting, these parts are called linked and if removable
called unlinked.  The mathematical classification and description of
such geometric or topological structures are called knot theory, and
it has a hundred year long history.

\begin{figure}
  \epsfig{file=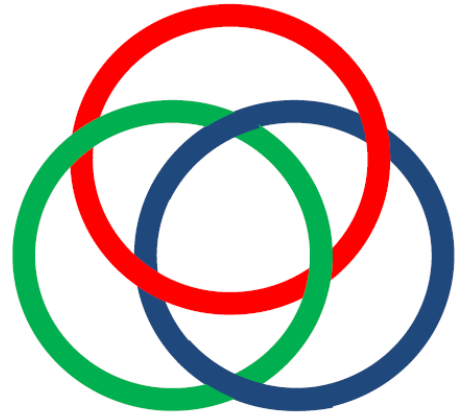,scale=0.6}
  \caption{Illustration of the Borromean rings which are held together
    only by the presence of all three. If any one of the rings is cut
    open, the entire structure falls apart.}
  \label{rings}
\end{figure}

One of the simplest cases is obviously three rings.  They can be
arranged so that they are linked, but if any one is cut and removed
the two remaining ones can be removed without further cutting. This is
illustrated in figure~\ref{rings}.  This construction is now often
called Borromean rings, since they are the heraldic symbol of the
dukes of Borromeo.  However, the structure was fascinating already in
the Norse mythology, shown as triangles and known as Odins knot, and
in various disguises especially in religious contexts throughout
history. Not surprisingly psychology (emphasized by for instance
Jacques Lacan) could use it a ambly as symbols of interlocked triads
like past present and future, etc.

In science, the Borromean rings have triggered several
generalizations.  The present paper discusses whether a new family of
generalized structures may have counterparts in physics. We shall
start by giving a brief review of previous work.  Several early papers
\cite{tai1876,bru1892} preceeded the first mathematically rigorous
treatments \cite{deb61,pen69} in the 1960's.  Since then links of
multiplicity $n$, essentially $n$ (deformed) rings in
three-dimensional space, are called Brunnian if they form a connected
structure where all sybsystems are unlinked.

A link is of type $B(n,k)$ if it consists of $n$ rings in such a way
that any subsystem of $m$ rings with $m \leq k$, is unlinked.  For $k
= 2$ the link is called Borromean, for $k = n - 1$ it is called
Brunnian.  If $n = 3$ the two notions coincide.  The notion of
$B(n,k)$ links was introduced in \cite{NewStates,Baas12} extending
previous notions in \cite{lia94}.

\begin{figure}
  \subfigure[A first order Brunnian ring]{
    \epsfig{file=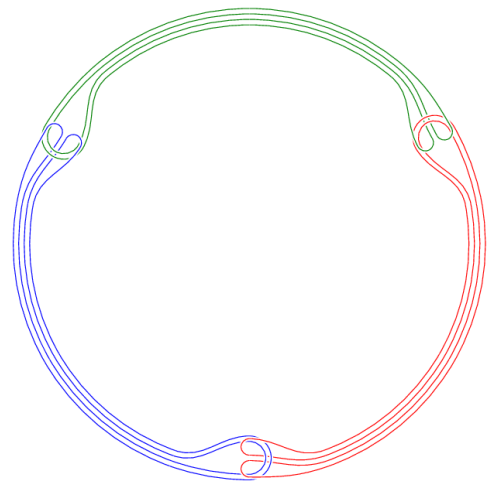,width=0.75\linewidth}
  } \qquad \subfigure[A second order Brunnian ring]{
    \epsfig{file=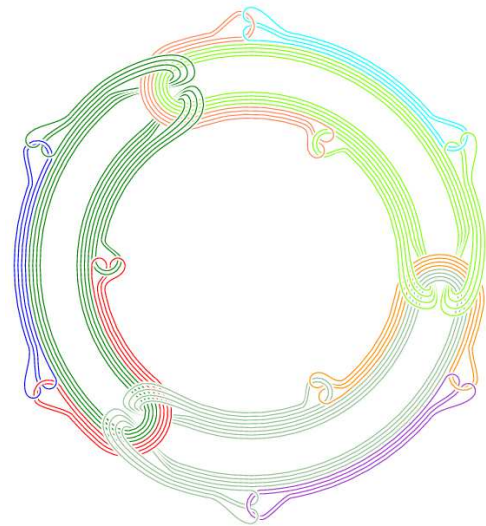,width=0.75\linewidth}
  }
  \caption{Brunnian rings of first and second order}
  \label{Brunnian}
\end{figure}

Further generalizations of higher order links --- in particular higher
order Brunnian links were made in \cite{NewStates} as special case of
higher order structures introduced in \cite{Baas09}, expanded in
\cite{Structure}.  First order structures are for example Brunnian
links, second order structures are Brunnian rings of Brunnian rings,
etc., see figures~\ref{rings}--\ref{cluster}.  See \cite{NewStates} for
more illustrations of higher order links.

We shall focus on the occurence on related higher-order structures in
physics, but should point out first that more direct examples may be
found in chemistry or biology. Molecular rings with non-standard
topologies have been contemplated for a long time \cite{fri61} and may
in certain cases have different properties (such as optical activity).
Biological molecules are known to form long chains of folded
structures, often with knots.  Borromean rings have been
constructed from DNA \cite{Mao97} and even high-order topological
structures have been envisaged \cite{Baas12}.

In \cite{NewStates} it was proposed that higher order links for
example of Brunnian type may suggest new physical states.  The concept
of Borromean systems was introduced in microscopic physics
\cite{Zhu92} based not on geometrical linking (topology), but on
binding energy. We shall in a similar manner take over all of the
above topological concepts by replacing geometrically (un)linked by
energetically (un)bound. Note that since the exact geometrical
configurations now are of less direct importance we can (and will)
also consider two-dimensional structure rather than confining
ourselves to three spatial dimensions. In the following section we
shall give a general discussion of the possibilities for having higher
order Brunnian systems. This is followed by more specific looks at
atomic, condensed matter and nuclear systems.

\section{Physical realizations of higher order Brunnian systems}
In order to realize higher-order Brunnian systems in physical setups,
one needs to consider multi-particle dynamics. For concreteness, we
will focus here on the case of a second-order Brunnian state built
from nine particles. At first we discuss the basic Hamiltonian setup
and the interactions that one can expect. We then go on to consider
subfields of physics where such a system might be accessible. This
includes ultracold atomic gases of both non-polar and polar single-
and two-species gases and nuclear systems. We also make some comments
about potential relevance of higher-order Brunnian structures in
traditional condensed-matter and solid-state systems.

\begin{figure}
  \epsfig{file=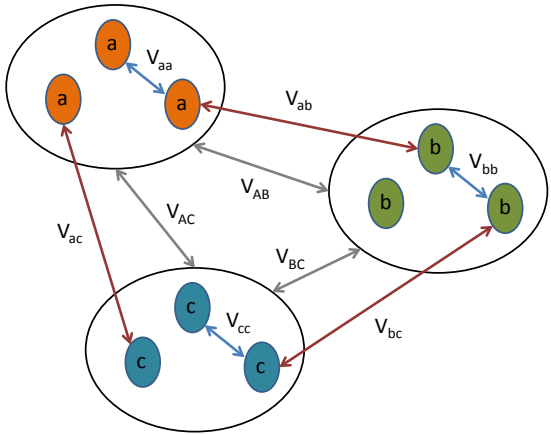,scale=0.33}
  \caption{A possible second-order Brunnian system with nine particles
    grouped into three different species, $a$, $b$, and $c$. The
    assumption is that each species holds a Borromean three-body bound
    state (indicated by the black rings). The intra-species
    interactions are denoted $V_{aa}$, $V_{bb}$, and $V_{cc}$, while
    the inter-species are $V_{ab}$, $V_{bc}$, and $V_{ac}$. In an
    effective description, we assume that we have three clusters that
    interact through effective interactions $V_{AB}$, $V_{BC}$, and
    $V_{AC}$.}
  \label{cluster}
\end{figure}

Assume that we have three species of particles, $a,b,c$, which could
be distinguished by either internal quantum numbers (f.x. electron
spin or hyperfine spin in neutral atoms) or external quantum numbers
(f.x. different mass particle). At this point we do not assume
anything about the quantum statistics of the different species and
they can as such be either fermionic, bosonic, or a mixture of the
two. We assume that the system has two-body interaction terms between
the same and different species, which we denote
$V_{aa},V_{bb},V_{cc},V_{ab},V_{bc},V_{ac}$ with the subscript
indicating the particular pair of particles. These interactions are
indicated on figure~\ref{cluster}.

In order to have a second-order Brunnian system of nine particles, we
need to have three Borromean three-body systems, that is three bound
three-body subsystems wherein no two-body interaction supports a bound
state. This is well-known to occur on the weakly-interacting side of a
Feshbach resonance in cold atomic gas systems that display the Efimov
effect \cite{coldefimov}. Numerically, such a state can be produced by
using model potentials such as Gaussian, or Yukawa type interactions
with interaction ranges that are short compared to the overall size of
the Borromean three-body states \cite{Jen04}. In this type of setup we
are in a sense working effectively with three Borromean systems and
ask when those three will form a bound structure (see
figure~\ref{cluster}).  The idea is then that the effective interaction
between these three-body Borromean clusters should support a Borromean
state, i.e. if you remove one cluster you break the system into three
unbound three-body states where each of the three-body systems form a
Borromean state.

The total potential is 
\begin{equation}
  V=V_{aa}+V_{bb}+V_{cc}+V_{ab}+V_{bc}+V_{ac},
\end{equation}
and the Schr{\"o}dinger equation in turn becomes
\begin{equation}
  H\Psi:=\left(T+V\right)\Psi=E\Psi,
\end{equation}
where $\Psi$ is the total (nine particle) wave function, $E$ is the
energy, and $T$ is the relative kinetic energy operator (excluding the
center of mass part). Pursuing the idea above that one should be able
to describe the system as a clusterized structure with three Borromean
three-body systems, we now restructure the Hamiltonian, $H$. Following
the schematics in figure~\ref{cluster}, we denote the interactions
between each of the clusters as $V_{AB}$, $V_{BC}$, and $V_{AC}$,
which generally depend on the two-body interactions with the lower
case subscripts, $V_{ij}$.  For a well-developed cluster structure it
would be a good approximation to decompose the total wave function as
$\Psi=\phi_A\phi_B\phi_C \Phi_{ABC}$, where $\phi_i$ ($i=A,B,C$)
satisfies
\begin{equation}
  \label{wavein}
  (T_i+V_i)\phi_i=E_i\phi_i,
\end{equation}
where $T_i$ is the relative kinetic energy operator of the
corresponding three-body system, and
$V_A=V_{aa}(12)+V_{aa}(23)+V_{aa}(13)$ denotes the potential of all
pairs within the three-body cluster containing species $a$, and
likewise for $B$ and $C$. The 'relative' wave function between the
three clusters, $\Phi_{ABC}$, satisfies the equation
\begin{equation}
  \label{wavecl}
  (T_{ABC}+V_{AB}+V_{BC}+V_{AC})\Phi_{ABC}=E_{ABC}\Phi_{ABC},
\end{equation}
where 
\begin{equation}
  T_{ABC}=-\frac{\hbar^2}{2M_A}\vec{\nabla}_{\vec{R}_{CM}^{A}} -
  \frac{\hbar^2}{2M_B}\vec{\nabla}_{\vec{R}_{CM}^{B}} - T_{CM} 
  -\frac{\hbar^2}{2M_C}\vec{\nabla}_{\vec{R}_{CM}^{C}},
\end{equation}
is the kinetic energy operator of the relative motion of the
clusters. Here we have introduced the total mass $M_A=3m_a$ and the
center-of-mass coordinate of the cluster $\vec{R}_{CM}^{A}$ (likewise
for $B$ and $C$).

In order to achieve a realization of a second-order Brunnian system,
one needs (i) to find a physical system where (\ref{wavecl})
supports a Borromean state, i.e. regime where no two-cluster subsystem
is bound but the total three-cluster system is bound, and (ii) the
subsystems governed by (\ref{wavein})  must also be Borromean,

\section{Ultracold Atomic Gases}
Advances in cold atomic gas physics \cite{bloch2008,lewenstein2008}
implies that these systems now provide an effective simulation venue
for testing various hypothesis of quantum mechanics, both for few- and
many-body systems. In particular, the control over interactions is
extremely precise through use of Feshbach resonances \cite{chin2010}
and facilitates access to the strongly-coupled regime where typical
perturbative approaches fail. Recently, cold polar molecules have been
added to the toolbox \cite{lahaye2009,ye2009}. Heteronuclear molecules
have a permanent dipole moment that is tunable and can provide a very
strong long-range and anisotropic interaction. This extends the
potential for simulation of quantum systems that are similar to
materials since dipolar interactions can mimic the behavior of systems
that have long-range Coulomb forces.

\subsection{Non-polar atomic gases}
First we consider the traditional case of non-polar atomic
gases. These are typically made from single-species alkali atoms, but
experiments have also been done with homonuclear molecules which, in
absence of an external aligning field, will display no polar
character.  At the low densities and the low temperatures of
experimentally realized atomic gases, the interactions that govern the
dynamics of the systems have very short-range \cite{bloch2008}.  Since
only low-energy scattering is allowed, this implies that the
effective-range expansion can be used to describe the effects of
interactions. With Feshbach resonances one can in fact tune the
scattering length, $a$, to any desired value \cite{chin2010}, and it
is this feature that we will exploit.

In order to implement the three-species situation above we first
consider bosonic alkali atoms and assume that we have populated three
internal (hyperfine) states. This has been realized in experiments,
effectively creating a spin-one system, i.e. bosons with three
internal degrees of freedom \cite{chapman,kurn}. What is important is
the particular scattering lengths in the different channels $a,b,c$.
However, this requires that one can find a set of Feshbach resonances
that will allow all intra-species scattering lengths to be tuned to
the negative $a$ side, and close enough to the Feshbach resonance
($|a|\to\infty$) where the Borromean bound state exists. At this point
one could perhaps bind the second-order Brunnian nine particle system,
perhaps with further tuning of the scattering lengths (without going
to the $a>0$ side where two-body subsystem become bound in any of the
three hyperfine components).  A problem with the species one usually
encounters for experiments with spin-one bosons like $^{87}$Rb, is
that the background scattering length is positive, i.e. there is a
two-body bound state between two atoms from the start. This means the
system does not fulfil our aim without further fine-tuning.

The scenario with internal hyperfine state of a single atom is
probably hard to achieve unless a fortunate set of resonances are
present, or perhaps with suitable combinations of magnetic and optical
resonances \cite{chin2010}. A different route would be to use
different atoms as the different species. This requires use of
inter-species Feshbach resonances instead. Again they should be
bosonic atoms so as to allow a Borromean state to form in each
cluster. Preferably, the intra-species interaction should have a
negative background value of the scattering length so that no two-body
bound state is present, in constrast to the $^{87}$Rb system discussed
above. Fortunately, there are some good alkali species available that
have just this feature such as $^{7}$Li, $^{39}$K, or $^{85}$Rb.  Thus
far, however, inter-species interactions, resonances, and background
parameters are largely unknown.

\subsection{Dipolar molecules}
We now proceed to discuss potential realization with polar molecules
that have permanent dipole moments. In this case, externally applied
AC and DC electric fields can be used to manipulate the interaction
into various interesting regimes \cite{micheli2007}. As discussed
above, we are interested in a regime of Borromean binding. Thus we
need to look for cases where two-body bound states are absent but
three-body trimers can be bound nonetheless. However, the catch with
dipolar molecules is that they can have very strong chemical losses
due to strong head-to-tail attraction \cite{lahaye2009,ye2009}.  This
can be suppressed by confining the molecules in low-dimensional
structures such as tubes or planes. This has been recently shown to
reduce chemical losses significantly \cite{miranda2011}.

\begin{figure}
  \epsfig{file=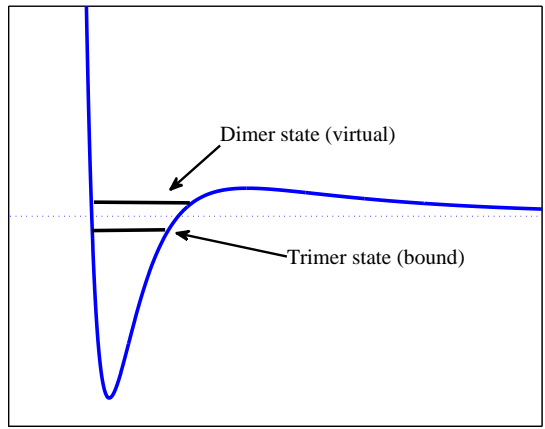,scale=0.45}
  \caption{Schematic illustration of a potential with an inner
    repulsive core, mid-range attractive pocket and an outer repulsive
    tail. The dotted horizontal line marks zero energy with respect to
    infinity where the potential vanishes. This potential shape can
    support a bound three-body trimer state even when the dimer state
    is unbound.}
  \label{potential}
\end{figure}

Interestingly, it is possible to create a two-dimensional setup and
apply external fields to produce potentials that allow Borromean
binding. In two dimensions, it is known that Borromean states can
occur in potentials that have an inner attractive pocket and an outer
repulsive tail \cite{nielsen97}. Such a potential can in fact be
achieved with polar molecules \cite{micheli2007,lee2012}. A schematic
depiction of this is shown in figure~\ref{potential}.  Three identical
fermionic particles interacting via such a potential will
support a bound trimer when no two-body subsystem is bound
\cite{artem2012}.

\begin{figure}
  \epsfig{file=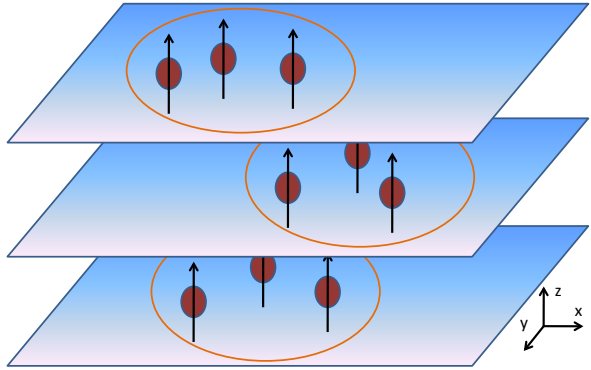,scale=0.33}
  \caption{A setup with three parallel planes containing polar
    molecules. An AC and DC electric field is applied to induce a
    specific potential between the dipoles in the planes and between
    the planes. The dipole moments are shown as arrows perpendicular
    to the planes. Through tuning of the potentials, one can achieve a
    situation where there are trimer bound states of three dipolar
    molecules in each plane. To achieve a second order Brunnian
    structure the total system across the planes would have to be
    bound such that removal of one trimer (or layer) makes the
    structure dissociate.}
  \label{trilayer}
\end{figure}

Having fulfilled the first condition of presence of Borromean
subsystems, we must now consider how to generate a nine particle three
cluster system. However, with polar molecules we have long-range
interactions. This implies that if we take a layered setup with three
planes each containing one Borromean cluster, we could potentially
achieve our goal of producing a second-order Brunnian structure. In
figure~\ref{trilayer} we show an illustration of a trilayer setup. By
adjusting the AC and DC external fields one tunes not only the
intra-plane but also the inter-plane interactions, and it is not
inconceivable that a regime can be found where the total structure of
nine polar molecules is bound in a Borromean fashion such that
removing one plane will break up the structure.

The DC field that is applied to the system will tend to align the
molecules perpendicular to the layer planes as indicated by the arrows
in figure~\ref{trilayer}. The interaction between particles in different
layers holds a two-body bound state in the case of no AC field
\cite{jeremy2010,artem2011a,artem2011b}.  However, the presence of an
AC field component will in general change the structure of the
potential and can be used to push the two-body bound state above
threshold \cite{micheli2007}. We expect that the perfect conditions to
look for a second-order Brunnian system would be somewhere around the
threshold but with the two-body state slightly above (rendering it
virtual).

In terms of experimental measurement, we imagine that one can use
traditional RF spectroscopic techniques to probe the bound state
structures \cite{RF}. The RF method works by driving transitions
between different internal hyperfine states in the molecules. If one
drives the molecules into a hyperfine state where the interaction
potentials are very different and will not provide a bound structure,
then the RF response pattern will display a structure at the binding
energy of the initial state. This has been used very recently to
detect Efimov states in association experiments
\cite{lompe2010,japanese2011}.

\section{Condensed-matter systems}
In general many-body systems such as those found in condensed-matter
and solid-state there are typically a hierachy of
correlations. Depending on structure and density, the most important
correlations tend to be those of relatively few particles. For
instance, two-body correlations will be very pronounced in relation to
weakly-interacting Bose-Einstein condensates \cite{ole2004}. However,
also three- and more-body correlations are important in such systems
\cite{truscott2011}. Recently, there has been a surge of activity into
the regime of strong interactions, which can be achieved for instance
by tuning parameters such that the scattering length diverges. Cold
atoms have provided an experimental route to the study of this
dynamics \cite{contact} which is considered a vital ingredient in
understanding high-temperature superconducting materials
\cite{hightc-reviews}. This strongly interacting regime in fact
provides a venue for disentangling the contributions of both two-
\cite{con2body1,con2body2} and three-body correlations
\cite{con3body1,con3body2,con3body3}, and in principle even
higher. This sort of direction seems a likely avenue for the study of
higher-order Brunnian systems.

Many-body correlations can typically have very non-local character,
such as in quantum and fractional quantum Hall systems
\cite{altland}. This is strongly connected to the topological
properties of these system which has recently been given a very
elegant description in momentum-space through the use of topological
invariants akin to the linking number or the genus of a surface
\cite{joelmoore2010}. The quantum Spin Hall effect and experimental
realizations of so-called topological insulators that have
topologically protected conducting surface states are recent
highlights of this research \cite{hasan2011,zhang2012}. The non-local
character of these interesting many-body phases could have some
relation to the higher-order Brunnian states that we discuss here.

The non-locality of the Borromean system can be attributed to the
defining feature that removing one particle will unbind the
system. This implies that no single particle, and no two-particle
combination, holds the entire information about the total system
structure. This is akin to the many synergetic relationships seen in
biological systems in Nature. Taking the Borromean system and
generalizing to higher-order Brunnian is a systematic extension of the
non-locality. Consider the example of the nine particle with three
clusters of three particle second-order Brunnian state discussed
previously. Here no single cluster contains the information that the
total structure is bound. Furthermore, removing one particle in any
one of the clusters will dissociate the clusters, and subsequently the
entire structure. This is truly a very non-local behavior similar to
the new topological systems intensely studied in condensed-matter and
solid-state systems.

A further, yet more speculative implication, is that generalized
Brunnian systems might be used to classify excitations in systems that
are typically described by conformal field theory such as quantum Hall
systems \cite{ezawa2008}. The correlations in the ground-state or
excitations could perhaps be described in terms of generalized
Brunnian structures embedded in the many-body environment.

\section{Nuclear systems}
We start by outlining the Brunnian systems of first order.  In
three-body Borromean (or Brunnian) systems in nuclei at least two out
of the three components must be neutrons, protons or alpha particles
in order to maintain the identity of the components in the combined
system. Apart from $^9$Be all established Borromean ground states
occur at the proton or neutron dripline and have a core+2p or core+2n
structure. Systems with several alpha particles are known, but appear
as excited states and even the simplest of them, the Hoyle state at
7.65 MeV in $^{12}$C, is unbound. Among more complex states, the
ground state of $^{123}$Xe fulfills the energy requirements for being
a Borromean state built of $^{115}$Sn and two alpha particles, but it
is not clear that a cluster description is appropriate in this case.
The core plus two nucleon systems naturally occur at the driplines due
to the requirement that the core plus one nucleon system is
unbound. In light nuclei where Coulomb barriers for protons are small
and the angular momenta are small as well, most Borromean systems
(whether in ground states or excited states) appear as two-neutron or
two-proton halo nuclei, see \cite{Jen04} that also contains a
discussion of the core+p+n systems.  Two Borromean halo nuclei we
shall discuss in the following are $^6$He and $^{11}$Li.  We mention
in passing that there is only one four-body Brunnian system known in
nuclei, namely $^{10}$C \cite{Cur08}.

The Brunnian systems of second order will by definition lie just below
the threshold for break-up into three three-body Brunnian systems of
first order. The energies of these thresholds are easily found and are
as follows: Three Hoyle states (we disregards for the moment that it
is slightly unbound) would give a second order state close to the 9
alpha threshold in $^{36}$Ar at 52.06 MeV. The three $^9$Be threshold
is at 48.63 MeV in $^{27}$Mg, the three $^6$He threshold is at 27.86
MeV in $^{18}$C, and the three $^{11}$Li threshold lies more than 50
MeV above the unbound ground state resonance in $^{33}$F.  That the
candidate states appear at high excitation energy is natural for
cluster states in nuclei as is known e.g.\ from the Ikeda picture
\cite{Ike68,Oer06} for alpha-cluster states in self-conjugate nuclei.

The candidate states for having a Brunnian structure of second order
would have a quite unusual spatial structure and would, as halo
nuclei, have a rather large extent. However, being so high up in the
continuum implies that there will be plenty of states with which they
can mix. They can therefore be expected to behave in close analogy to
halo states at high excitation energy that are known \cite{Jen00} to
mix with other states rather than to remain unperturbed. In both
cases, there is no quantum number or symmetry that preserves the
topology or clustering of the structure. The Brunnian structure has to
appear ``dynamically'' and cannot be imposed.

Even without the mixing with other states one may ask whether a
Brunnian structure of second order could appear at all. The point is
that we in nuclei have only a limited number of building blocks for
the Brunnian systems of first order. The interactions between the
components of different first order systems can therefore not be
neglected and will typically be of the same order as the ones within a
given first order system.  This will lead to distortions of the first
order systems unless the binding energy of the three first order
systems in the second order state is extremely low. In other words
there is no natural manner of separating the scales to make the
appearance of the second order Brunnian structure realistic.

Brunnian states of second order are therefore unlikely to be found in
nuclei. Still, related structures may appear and have been considered
earlier. Cluster systems in nuclei \cite{Oer06,Fre07} have been
discussed for many years and searches have been made e.g.\ for states
in $^{12}$Be with a ``two $^6$He structure''. This particular
combination may be favoured due to a threshold of only 10.1 MeV which
is low compared to other combinations of two first order Brunnian
nuclei.  Among the other cluster structures discussed are three-center
nuclear molecules in heavy carbon isotopes and also non-standard
topologies such as Wilkinson's suggestion of a ring of alpha-particles
bound by delocalized neutrons \cite{Wil86}, but these structures do
not involve a Brunnian aspect.  For more general topological cluster
type many body systems, see \cite{Structure}.

A special case is the addition of neutron pairs to an existing
nucleus. The (as yet unreachable) neutron dripline for heavy nuclei is
predicted to have many nuclei analogous to $^8$He and $^{19}$B, i.e.\
where the core+$x$n system is unbound for $x=1,3,5,..$ and bound for
$x=2,4$ (and even higher). It has even been suggested that this type
of structure may appear also for low masses beyond the dripline
\cite{Jen92}. However, one would expect the internal structure of the
states to rearrange as neutrons are added. For the only case that has
been studied in detail today, $^8$He, the structure is much better
described by four neutrons around an alpha-particle rather than two
di-neutrons around an alpha-particle.  Structurally these correspond
to Borromean systems with more than three components rather than
Brunnian systems.

\section{Discussion and Conclusion} 
Mathematical knot theory employs purely topological concepts.  This
mature discipline was over the years successively generalized to
classify and describe geometric structures of any number of rings.
About 30 years ago Borromean structure of nuclear clusters was found
to exhibit unexpected large spatial extension. This initiated huge
research efforts involving large-scale radioactive beam facilities
with the aim of localizing and investigating such nuclear structures.
This quickly gave rise to the very notion of Borromean systems as
bound three-body systems where all three two-body subsystems are
unbound.  Thus, the Borromean nuclear three-body system was defined in
terms of binding arising from interactions bewteen the particles
forming the three-body system.  The geometric aspect has disappeared
from this definition although the basic concept of entangled clusters
remains. The same generalization to more particles can then be done,
based on binding energies, precisely as for geometric structures.

The definitions can be based on the concept of a link, that is two
systems are linked if they are bound or if they can be detached from
each other without cutting one of the rings. Correspondingly they are
unlinked if they are unbound and hence would fall apart
instantaneously.  An $(n,k)$ Brunnian system is defined as a system
with $n$ particles where all subsystems with $m (\leq k)$ particles
are unlinked.

Recently surfaced another generalization to higher order Brunnian
systems.  We shall confine ourselves to $k=n-1$ and specify the
corresponding generalization to the next order for a system of nine
particles divided into three clusters forming a Borromean system,
where each cluster is a three-body Borromean system.  The original
topological picture suggests macroscopic systems which easily can be
visualized as configurations of a number of entangled rings.  It is
not unlikely that such structures occur in macroscopic biomolecules
consisting of a large number of atoms of molecules.  We shall not
investigate these possibilities but instead focus on the world of
microscopic quantum physics. 

We first specify the hamiltonian in general terms to relate to
descriptions in microscopic quantum physics. We discuss three
subfields of physics where such structures could be found.  The
interactions in nuclear physics is first of all of short range but
modifcations are essentially always present due to the long-range
repulsive Coulomb interaction.  This implies that a bound nuclear
structure must have a radius sufficiently small to exploit the
short-range attraction. For $s$-waves and neutral particles this is
not a strong limitation, but unfortunately all nuclear clusters
(except the neutron itself) are charged. The combination of
short-range attraction and the repulsive Coulomb force allows two
options, that is either be inside or outside the resulting barrier.
Being outside means unbound and being inside the barrier implies a
small radius.  Therefore, a bound system almost inevitably has a
relatively small distance between all particles. The necessary cluster
configurations ($3 \times 3$) is then very unlikely in a ground (or
low-lying excited) state, simply because the interactions between any
pair of particles (within and between three-body subcluster particles)
are comparable.

Cold atomic gases are presently investigated in many laboratories with
entirely new techniques.  It is possible to vary the interaction
between pairs of particles by using couplings between excited states
depending on externally tunable magnetic fields. By combining with
laboratory controlled electromagnetic fields from lasers, both
geometries and effective interactions can be varied to a large extent.
These systems are therefore tempting candidates for higher order
microscopic Brunnian systems.  First, three different states must be
available becasue all particles in one state, as bosons are allowed to
do, would not be able to produce a three-cluster configuration, but
rather a condensate. Still, this is possible and achieved, and each
state can be prepared to avoid two-body binding while allowing
Borromean structure. Then four-body and higher cluster correlations
should not be formed with a suitably low density, but the interactions
between the three different Borromean systems now also must be of
Borromean nature.  This requires experimental investigation and
probably a good deal of luck since these interactions cannot be tuned
independently.  

Another possibility for cold atomic gases is to look for
two-dimensional structures where the different Borromean systems
should be made in separate planes. The interactions are well-known for
polarized molecules in controlled electric fields. It is then possible
to apply different fields to tune the interactions between molecules
in separate planes.  We can then imagine to achieve the desired
Borromean conditions both within and between planes. 

Condensed matter systems can be of very complex structure. Whether
higher order Brunnian systems may be designed or just turn up is
difficult to say but far from excluded. However, we have no specific
knowledge of structures that suggest themselves in this context, at
best only very speculative in connection with non-local interactions
in recently studied quantum systems.

One difficulty is that spatially extended clusters are less likely
when the number of clusters is larger than three.  If several clusters
are bound together by short-range interactions, a barrier is
efffectively produced. Its effect is similar (less pronounced) to that
of the repulsive long-range Coulomb interaction which either pushes to
no binding or to spatially small sizes where the cluster structure is
difficult to maintain.
 
In conclusion, higher order Brunnian systems are, unlikely to occur in
nuclei, possible for cold atomic gases either for polarized molecules
in two dimensions or populations of three different states in three
dimensions, imaginable in condensed matter systems but may be
difficult to find or synthesize.

\end{document}